
\magnification=1200
\baselineskip=6mm
{\nopagenumbers
\noindent
\null\vskip 1truecm
\centerline
{\bf WIGNER DISTRIBUTION FUNCTION}
\centerline
{\bf FOR THE HARMONIC OSCILLATOR WITH DISSIPATION}
\centerline
{\bf WITHIN THE THEORY OF OPEN QUANTUM SYSTEMS}
\vskip 1truecm
\centerline {A. Isar}
\centerline {\it Department of Theoretical Physics, Institute of Atomic
Physics}
\centerline {POB MG-6, Bucharest, Romania}
\vskip 1truecm
\centerline {ABSTRACT}
\vskip 0.5truecm

Time evolution of the expectation values of various dynamical operators of the
harmonic oscillator with dissipation is analitically obtained within the
framework of the Lindblad's theory for open quantum systems. We deduce the
density matrix of the damped harmonic oscillator from the solution of the
Fokker-Planck equation for the coherent state representation, obtained from the
master equation for the density operator. The Fokker-Planck equation for the
Wigner distribution function is also solved by using the Wang-Uhlenbeck method
of transforming it into a linearized partial differential equation for the
Wigner function, subject to either the Gaussian type or the $\delta$-function
type of initial conditions. The Wigner functions which we obtain are
two-dimensional Gaussians with different widths.

\vfill\eject
}
\pageno=1
{\bf 1. Introduction}

In the last two decades, the problem of dissipation in quantum mechanics, i.e.
the consistent description of open quantum systems, was investigated by various
authors [1-6]. Because dissipative processes imply irreversibility and,
therefore, a preferred direction in time, it is generally thought that quantum
dynamical semigroups are the basic tools to introduce dissipation in quantum
mechanics. The most general form of the generators of such semigroups was given
by Lindblad [7-9]. This formalism has been studied for the case of damped
harmonic oscillators [8,10,11] and applied to various physical phenomena, for
instance, the damping of collective modes in deep inelastic collisions in
nuclear physics [12-14] and the interaction of a two-level atom with the
electromagnetic field [15].

This paper, also dealing with the damping of the harmonic oscillator within the
Lindblad theory for open quantum systems, is concerned with the time evolution
of various dynamical operators involved in the master and Fokker-Planck
equations, in particular with the time development of the density matrix. In
[14], by using the characteristic function of the density operator, we obtained
analytical expressions for the first two moments of coordinate and momentum of
the harmonic oscillator with dissipation. In [16] the Lindblad master equation
was transformed into Fokker-Planck equations for quasiprobability distributions
and a comparative study was made for the Glauber $P$, antinormal ordering $Q$
and Wigner $W$ representations. In [17] the density matrix of the damped
harmonic oscillator was  represented by a generating function, which was then
obtained by solving a time-dependent linear partial differential equation
derived from the master equation.

The aim of this work is to explore the physical aspects of the Fokker-Planck
equation which is the $c$-number equivalent equation to the Lindblad master
equation. Generally the master equation for the density operator gains
considerably in clarity if it is represented in terms of the Wigner
distribution function which satisfies the Fokker-Planck equation. It is worth
mentioning that the master and Fokker-Planck equations agree in form with the
corresponding equations formulated in quantum optics [18-25].

The content of this paper is arranged as follows. In Sec.2 we review briefly
the derivation of the master equation for the density operator of the harmonic
oscillator. In order to get an insight into physical meanings of the master
equation, we first split it up into several equations satisfied by the
expectation values of the dynamical operators involved in the master equation.
These equations are then solved analytically. In Sec.3 we transform the master
equation into the Fokker-Planck equation by means of the well-known methods
[4,26-30]. We show how the density matrix, subject to the Poisson distribution
type initial condition, is extracted with the help of the solution of the
Fokker-Planck equation for the coherent state representation. Then the
Fokker-Planck equation for the Wigner distribution is solved by the
Wang-Uhlenbeck method of transforming it into a linearized partial differential
equation for the Wigner function, subject to either the Gaussian type or the
$\delta$-function type of initial conditions. Finally, conclusions are given
in Sec.4.

{\bf 2. The master equation for the damped quantum harmonic oscillator}

The rigorous formulation for introducing the dissipation into a quantum
mechanical system is that of quantum dynamical semigroups [2,3,7]. According to
the axiomatic theory of Lindblad [7,9], the usual von Neumann-Liouville
equation ruling the time evolution of closed quantum systems is replaced in the
case of open systems by the following equation for the density operator $\rho$:
$${d\Phi_{t}(\rho)\over dt}=L(\Phi_{t}(\rho)).\eqno (2.1)$$
Here, $\Phi_{t}$ denotes the dynamical semigroup describing the irreversible
time evolution of the open system in the Schr\"odinger representation and $L$
the infinitesimal generator of the dynamical semigroup $\Phi_t$. Using the
structural theorem of Lindblad [7] which gives the most general form of the
bounded, completely dissipative Liouville operator $L$, we obtain the explicit
form of the most general time-homogeneous quantum mechanical Markovian master
equation:
$${d\rho(t)\over dt}=L(\rho(t)),\eqno (2.2)$$
where
$$L(\rho(t))=-{i\over\hbar}[H,\rho(t)]+{1\over 2\hbar}\sum_{j}([V_{j}
\rho(t),V_{j}^+]+[V_{j},\rho(t)V_{j}^+]).\eqno (2.3)$$
Here $H$ is the Hamiltonian of the system. The operators $V_{j}$ and $V_{j}^+$
are bounded operators on the Hilbert space $\bf H$ of the Hamiltonian.

We should like to mention that the Markovian master equations found in the
literature are of this form after some rearrangement of terms, even for
unbounded Liouville operators. In this connection we assume that the general
form of the master equation given by (2.2), (2.3) is also valid for unbounded
Liouville operators.

In this paper we impose a simple condition to the operators $H,V_{j},V_{j}^+$
that they are functions of the basic observables $q$ and $p$ of the
one-dimensional quantum mechanical system (with $[q,p]=i\hbar$) of such kind
that the obtained model is exactly solvable. A precise version for this last
condition is that linear spaces spanned by first degree (respectively second
degree) noncommutative polynomials in $p$ and $q$ are invariant to the action
of the completely dissipative mapping $L$. This condition implies [8] that $V_
{j}$ are at most first degree polynomials in $p$ and $q$ and $H$ is at most a
second degree polynomial in $p$ and $q$. Then the harmonic oscillator
Hamiltonian $H$ is chosen of the form
$$H=H_{0}+{\mu \over 2}(pq+qp),~~~H_{0}={1\over 2m}p^2+{m\omega^2
\over 2}q^2.\eqno (2.4)$$
With these choices the Markovian master equation can be written [10]:
$${d\rho \over dt}=-{i\over \hbar}[H_{0},\rho]-{i\over 2\hbar}(\lambda +\mu)
[q,\rho p+p\rho]+{i\over 2\hbar}(\lambda -\mu)[p,\rho q+q\rho]$$
$$-{D_{pp}\over {\hbar}^2}[q,[q,\rho]]-{D_{qq}\over {\hbar}^2}[p,[p,\rho]]+
{D_{pq}\over {\hbar}^2}([q,[p,\rho]]+[p,[q,\rho]]),\eqno (2.5)$$
where $D_{pp},D_{qq}$ and $D_{pq}$ are the diffusion coefficients and $\lambda$
the friction constant. They satisfy the following fundamental constraints [10]:
$${\rm i})~D_{pp}>0,~~{\rm ii})~D_{qq}>0,~~{\rm iii})~D_{pp}D_{qq}-D_{pq}^2\ge
{\lambda}^2{\hbar}^2/4.\eqno(2.6)$$
The equality $\mu=\lambda$ is a necessary and sufficient condition for $L$ to
be translation invariant [8]: $[p,L(\rho)]=L([p,\rho]).$ In the following
general values for $\lambda$ and $\mu$ will be considered. In the
particular case when the asymptotic state is a Gibbs state
$$\rho_G(\infty)=e^{-{H_0\over kT}}/{\rm Tr}e^{-{H_0\over kT}},\eqno(2.7)$$
these coefficients reduce to
$$D_{pp}={\lambda+\mu\over 2}\hbar m\omega\coth{\hbar\omega\over 2kT},
{}~~D_{qq}={\lambda-\mu\over 2}{\hbar\over m\omega}\coth{\hbar\omega\over 2kT},
{}~~D_{pq}=0,\eqno(2.8)$$
where $T$ is the temperature of the thermal bath.

Introducing the annihilation and creation operators via the relations
$$q=\sqrt{{\hbar\over 2m\omega}}(a^++a), ~~p=i\sqrt{{\hbar m\omega\over 2}}
(a^+-a),\eqno(2.9)$$
we have $H_{0}=\hbar\omega(a^+a+{1\over 2})$ and the master equation (2.5)
takes the form
$${d\rho\over dt}={1\over 2}(D_{1}+\mu)(a^+ a^+\rho-a^+\rho a^+)+{1\over 2}
(D_{1}-\mu)(\rho a^+ a^+-a^+\rho a^+)$$
$$+{1\over2}(D_{2}+\lambda+i\omega)(a\rho a^+-a^+ a\rho)+{1\over 2}(D_{2}-
\lambda-i\omega)(a^+\rho a-\rho aa^+)+{\rm H.c.},\eqno (2.10)$$
where
$$D_{1}={1\over\hbar}(m\omega D_{qq}-{D_{pp}\over m\omega}+2iD_{pq}),
{}~~D_{2}={1\over\hbar}(m\omega D_{qq}+{D_{pp}\over m\omega}). \eqno (2.11)$$

In the literature, master equations of the type (2.5) or (2.10) are encountered
in concrete theoretical models for the description of different physical
phenomena in quantum optics [18-24], in treatments of the damping of collective
modes in deep inelastic collisions of heavy ions [12-14,31-34] or in the
quantum mechanical description of the dissipation for the one-dimensional
harmonic oscillator [4,6,8,10,28,29]. A classification of these equations,
whether they satisfy or not the fundamental constraints (2.6), was given in
[17].

The meaning of the master equation becomes clear when we transform it into
equations satisfied by various expectation values of operators involved in the
master equation, $<A>={\rm Tr}[\rho(t)A],$ where $A$ is an operator composed
of the creation and annihilation operators. Multiplying both sides of (2.10)
by $a$ and taking throughout the trace, we get
$${d\over dt}<a>=-(\lambda+i\omega)<a>+\mu<a^+>.\eqno(2.12)$$
Similarly, the equation for $<a^+>$ is given by
$${d\over dt}<a^+>=-(\lambda-i\omega)<a^+>+\mu<a>.\eqno(2.13)$$
In the absence of the second term on the right-hand side of (2.12) and (2.13),
the two equations represent independently a simple equation of oscillation
with damping. By coupling (2.12) to (2.13) we get a second order differential
equation for either $<a>$ or $<a^+>$. For example, we obtain:
$${d^2\over dt^2}<a>+2\lambda{d\over dt}<a>+(\lambda^2+\omega^2-\mu^2)<a>=0,
\eqno(2.14)$$
which is the equation of motion for Brownian motion of a classical oscillator,
but without the term corresponding to random process. Because of the vanishing
terms on the right-hand side, we may equally state that (2.14) is the equation
of motion with zero expectation value of the random process. In the study of
the Brownian motion of a classical oscillator, one replaces the second-order
differential equation of motion of the type (2.14) by two equivalent
first-order differential equations [35] which are precisely the Langevin
equations. Accordingly, we may say that (2.12) and (2.13) are the Langevin
equations corresponding to (2.14), but without the random process term. The
integration of (2.14) is straightforward. There are two cases: a)$\mu>\omega$
(overdamped) and b)$\mu<\omega$ (underdamped). In the case a) with the notation
$\nu^2=\mu^2-\omega^2,$ we obtain:
$$<a(t)>= e^{-\lambda t}[<a(0)>(\cosh\nu t-i{\omega\over\nu}\sinh\nu t)+
{\mu\over\nu}<a^+(0)>\sinh\nu t].\eqno(2.15 {\rm a})$$
In the case b) with the notation $\Omega^2=\omega^2-\mu^2$, we obtain:
$$<a(t)>=e^{-\lambda t}[<a(0)>(\cos\Omega t-i{\omega\over\Omega}\sin\Omega t)
+{\mu\over\Omega}<a^+(0)>\sin\Omega t].\eqno(2.15 {\rm b})$$
Here $<a(0)>$ or  $<a^+(0)>$ stands for the initial value. The expression for
$<a^+(t)>$ can be obtained simply by taking the complex conjugate of the
right-hand side of (2.15).

For the computation of quantal fluctuations of the coordinate and momentum of
the harmonic oscillator, we need the expectation values of product operators,
such as $a^{+2}, a^2$ or $a^+a$. The dynamical behaviour of these product
operators can be well surveyed by deriving the equations satisfied by their
expectation values. By following the same procedure as before, employed in the
derivation of (2.12), we find:
$${d\over dt}<a^2>+2(\lambda+i\omega)<a^2>=2\mu<a^+a>+D_1+\mu,\eqno(2.16)$$
$${d\over dt}<a^{+2}>+2(\lambda-i\omega)<a^{+2}>=2\mu<a^+a>+D_1^*+\mu,
\eqno(2.17)$$
$${d\over dt}<a^+a>+2\lambda<a^+a>=\mu(<a^{+2}>+<a^2>)+D_2-\lambda.\eqno(2.18)
$$
Eqs.(2.16)-(2.18) exibit clearly that the expectation value of one double
operator has effects upon the dynamical behaviour of other double operators due
to the presence of non-zero terms on the right-hand sides. The solutions of
these equations are readily obtained by transforming them into two differential
equations satisfied by the sum and the difference of two double operators,
$<a^2>$ and $<a^{+2}>$. We obtain:
$${d^2\over dt^2}(<a^2>+<a^{+2}>)+4\lambda{d\over dt}(<a^2>+<a^{+2}>)+4(\lambda
^2+\omega^2-\mu^2)(<a^2>+<a^{+2}>)$$
$$={4\over\hbar}[(\lambda+\mu)m\omega D_{qq}-(\lambda-\mu){D_{pp}\over m\omega}
+2\omega D_{pq}]\equiv D,\eqno(2.19)$$
$${d\over dt}(<a^2>-<a^{+2}>)+2\lambda(<a^2>-<a^{+2}>)+2i\omega(<a^2>+<a^{+2}>)
=4i{D_{pq}\over\hbar}.\eqno(2.20)$$
The solution of (2.19) is straightforward and with the help of which both
equations (2.18) and (2.20) can be immediately solved. We find:
$$<a^2>=e^{-2\lambda t}[(1-i{\omega\over\nu})C_1 e^{2\nu t}+(1+i{\omega\over\nu
})C_2 e^{-2\nu t}-i{\mu\over\omega}C_3]+{D(\lambda-i\omega)\over 8\lambda
(\lambda^2-\nu^2)}+i{D_{pq}\over\hbar\lambda},\eqno(2.21 {\rm a})$$
$$<a^+a>=e^{-2\lambda t}[{\mu\over\nu}(C_1 e^{2\nu t}-C_2 e^{-2\nu t})+C_3]+{D
\mu\over 8\lambda(\lambda^2-\nu^2)}+{D_2-\lambda\over 2\lambda},\eqno(2.22 {
\rm a})$$
for the overdamped case $\mu>\omega$ and
$$<a^2>=e^{-2\lambda t}[(C_1+iC_2{\omega\over\Omega})\cos 2\Omega t+(C_2-iC_1{
\omega\over\Omega})\sin 2\Omega t-i{\mu\over\omega}C_3]+{D(\lambda-i\omega)
\over 8\lambda(\lambda^2+\Omega^2)}+i{D_{pq}\over\hbar\lambda},\eqno(2.21 {
\rm b})$$
$$<a^+a>=e^{-2\lambda t}[{\mu\over\Omega}(C_1\sin 2\Omega t-C_2\cos 2\Omega t)+
C_3]+{D\mu\over 8\lambda(\lambda^2+\Omega^2)}+{D_2-\lambda\over 2\lambda},\eqno
(2.22 {\rm b})$$
for the underdamped case $\omega>\mu$. The expression for $<a^{+2}>$ can be
obtained by taking the complex conjugates of (2.21{\rm a}) and (2.21{\rm b}).
Here, $C_1,C_2,C_3$ are the integral constants depending on the initial
expectation values of the operators under consideration. In particular, if
$D_{qq}=D_{pq}=0$ and $\mu=\lambda$ we obtain for the underdamped case $\omega
>\mu$ the same equations as those written by Jang in [31] for his model on
nuclear dynamics based on the second RPA at finite temperature. If the
condition $\mu\ll\omega$ is satisfied throughout in the underdamped case, the
expectation values (2.21{\rm b}) and (2.22{\rm b}) reduce to $(C_1+iC_2)\exp
[-2(\lambda+i\omega)t]+{D_1\over 2(\lambda^2+\omega^2)}(\lambda-i\omega)$ and
$C_3\exp(-2\lambda t)+{D_2-\lambda\over 2\lambda}$, respectively. These are the
underdamped solutions of (2.16) and (2.18), but in the right-hand sides
setting $\mu=0.$ For time $t\to\infty,$ we see from (2.22) that
$$<a^+a>={D\mu\over 8\lambda(\lambda^2+\omega^2-\mu^2)}+{D_2-\lambda\over 2
\lambda}.\eqno(2.23)$$
In the particular case when the asymptotic state is a Gibbs state (2.7), (2.8)
and $\mu=\lambda$, we get
$$<a^+a>={1\over 2}(\coth{\hbar\omega\over 2kT}-1)=(\exp{\hbar\omega\over
kT}-1)^{-1}\equiv <n>,\eqno(2.24)$$
which is the Bose distribution. This means that the expectation value of the
number operator goes to the average thermal-phonon number at infinity of time.
{}From the identity
$$<a^+a>=\sum_{m=0}^\infty m<m\vert\rho(t)\vert m>\eqno(2.25)$$
it follows
$$<m\vert\rho(\infty)\vert m>={<n>^m\over (1+<n>)^{m+1}}.\eqno(2.26)$$
In deriving this formula, we have made use of the identity $\sum_{m=0}^\infty
mx^m=x/(1-x)^2.$ The expression (2.26) shows that in the considered particular
case the density matrix reaches its thermal equilibrium -- the Bose-Einstein
distribution, whatever the initial distribution of the density matrix may be.
When the initial density matrix $<m\vert\rho(0)\vert m>$ is represented by a
distribution of the form $N^m/(1+N)^{m+1}$, where $N$ stands for the average
phonon number, the relation (2.25) implies that $<a^+a>=N.$ When the initial
density matrix is characterized by a distribution of the form
$${1\over m!}N^m e^{-N},\eqno(2.27)$$
(2.25) implies that $<a^+a>$ becomes also $N.$ Eq.(2.27) is nothing but a
Poisson distribution. If the initial density matrix is represented by a
Kronecker delta $\delta_{ms},$ we see from (2.25) that $<a^+a>=s,$ which
corresponds to the initial $s$-phonon state.

For the master equation (2.10) of the harmonic oscillator, physical observables
can be obtained from the expectation values of polynomials of the annihilation
and creation operators. So, for the position and momentum operators $q$ and $p$
via the relations (2.9), we can evaluate either the second moments or variances
(fluctuations), by making use of the results (2.15),(2.21),(2.22). Such
quantities have been also calculated [10,14], by using other methods.

{\bf 3. Fokker-Planck Equations}

One useful way to study the consequences of the master equation (2.10) for the
density operator of the one-dimensional damped harmonic oscillator is to
transform it into more familiar forms, such as the equation of motion for the
density matrix [17] or the equations for the $c$-number quasiprobability
distributions Glauber $P$, antinormal ordering $Q$ and Wigner $W$ associated
with the density operator [16]. In the second case the resulting differential
equations of the Fokker-Planck type for the distribution functions can be
solved by standard methods [26-29] employed in quantum optics and observables
directly calculated as correlations of these distribution functions.

{\it 3.1 Calculation of the density matrix from the Fokker-Planck equation}

The Fokker-Planck equation, obtained from the master equation and satisfied by
the Wigner distribution function $W(\alpha,\alpha^*,t)$, where $\alpha$ is a
complex variable, has the form [16]:
$${\partial W(\alpha,\alpha^*,t)\over\partial t}=-\{{\partial\over\partial
\alpha}[-(\lambda+i\omega)\alpha+\mu\alpha^*]+{\partial\over\partial\alpha^*}
[-(\lambda-i\omega)\alpha^*+\mu\alpha]\}W(\alpha,\alpha^*,t)$$
$$+{1\over 2}(D_1{\partial^2\over\partial\alpha^2}+D_1^*{\partial^2\over
\partial\alpha^{*2}}+2D_2{\partial^2\over\partial\alpha\partial\alpha^*})W(
\alpha,\alpha^*,t).\eqno(3.1)$$
When we substitute the $P$ representation function $P(\alpha,\alpha^*,t)$ for
$W(\alpha,\alpha^*,t)$ and the coefficients $D_1+\mu$ for $D_1, D_2-\lambda$
for $D_2$ in the above equation, we get the Fokker-Planck equation for the
coherent representation [16]. Despite the formal similarity to averaging with a
classical probability distribution, the function $P(\alpha,\alpha^*,t)$ is not
a true probability distribution. Because of the overcompleteness of the
coherent states, the $P$ function is not a unique, well-behaved positive
function for all density operators. Cahill [36] studied the $P$ representation
for density operators which represent pure states and found a narrow class of
states for which the $P$ representation exists. They can be generated from a
particular coherent state $\vert\alpha>$ by the application of a finite number
of creation operators. Also Cahill [37] introduced a representation of the
density operator of the electromagnetic field that is suitable for all density
operators and that reduces to the coherent state $P$ representation when the
latter exists. The representation has no singularities.

The Fokker-Planck equation (3.1) or the similar equation for the $P$
representation, subject to the initial condition
$$P(\alpha,\alpha^*,0)=\delta(\alpha-\alpha_0)\delta(\alpha^*-\alpha^*_0),\eqno
(3.2)$$
where $\alpha_0$ is the initial value of $\alpha$ can be solved [30,38,39] and
the solution to (3.1) (Green function) for the $P$ representation is found to
be
$$P(\alpha,\alpha^*,t)={2\over \pi\sqrt{{\rm det}\sigma(t)}}\exp\{-{1\over
{\rm
det}\sigma(t)}[\sigma_{22}(\alpha-\bar\alpha_0)^2+\sigma_{11}(\alpha^*-\bar
\alpha_0^*)^2-2\sigma_{12}\vert\alpha-\bar\alpha_0\vert^2\},\eqno(3.3)$$
where $\sigma\equiv(\sigma_{ij}),$
$$\sigma_{ij}(t)=\sum_{s,r=1,2}[\delta_{is}\delta_{jr}-b_{is}(t)b_{jr}(t)]
\sigma_{sr}(\infty).\eqno(3.4)$$
The function $\bar\alpha_0$ and its complex conjugate, which are still
functions of time, are given by
$$\bar\alpha_0=b_{11}(t)\alpha_0+b_{12}(t)\alpha_0^*.\eqno(3.5)$$
The functions $b_{ij}$ obey the equations
$$\dot b_{is}=\sum_{j=1,2}c_{ij}b_{js}\eqno(3.6)$$
with the initial conditions $b_{js}(0)=\delta_{js}$ and $\sigma(\infty)$ is
determined by
$$C\sigma(\infty)+\sigma(\infty)C^{\rm T}=Q^P,\eqno(3.7)$$
where
$$C=\left(\matrix{\lambda+i\omega&-\mu\cr
-\mu&\lambda-i\omega\cr}\right),~~~
Q^P=\left(\matrix{D_1+\mu&D_2-\lambda\cr
D_2-\lambda&D_1^*+\mu\cr}\right).\eqno(3.8)$$
We get
$$b_{11}=b_{22}^*=e^{-\lambda t}(\cos\Omega t-i{\omega\over\Omega}\sin\Omega t)
,~~b_{12}=b_{21}={\mu\over\Omega}e^{-\lambda t}\sin\Omega t,\eqno(3.9)$$
with $\Omega^2=\omega^2-\mu^2.$ While the functions $\sigma_{11},\sigma_{22}$
and $\bar\alpha_0$ are complex with $\sigma_{11}=\sigma_{22}^*$, the functions
${\rm det}\sigma(t)$ and $\sigma_{12}$ are real.

The solution of the Fokker-Planck equation has been written down providing the
diffusion matrix $Q^P$ is positive definite. However, the diffusion matrix in
the Glauber $P$ representation is not, in general, positive definite. If the
$P$ distribution does not exist as a well-behaved function, the so-called
generalized $P$ distributions can be taken that are well-behaved, normal
ordering functions [40]. The generalized $P$ distributions are nondiagonal
expansions of the density operator in terms of coherent states projection
operators.

In the coherent representation [41,42] the density operator $\rho(t)$ is
expressed by
$$\rho(t)=\int P(\alpha,\alpha^*,t)\vert\alpha><\alpha\vert d^2\alpha,\eqno
(3.10)$$
where $d^2\alpha=d({\rm Re}\alpha)d({\rm Im}\alpha)$ and $\vert\alpha>$ is the
coherent state. The matrix element of $\rho(t)$ in the $n$ quantum number
representation is obtained by multiplying (3.10) on the left by $<m\vert$ and
on the right by $\vert n>.$ By making use of the well-known relation
$$\vert\alpha>=\exp(-{1\over 2}\vert\alpha\vert^2)\sum_{n=0}^\infty{\alpha^n
\over\sqrt{n!}}\vert n>,\eqno(3.11)$$
we get
$$<m\vert\rho(t)\vert n>={1\over\sqrt{m!n!}}\int\alpha^n\alpha^{*m}P(\alpha,
\alpha^*,t)\exp(-\vert\alpha\vert^2)d^2\alpha.\eqno(3.12)$$
Upon introducing the explicit form (3.3) for $P(\alpha,\alpha^*,t)$ into
(3.12), we obtain the desired density matrix for the initial coherent state.
However, due to the powers of complex variables $\alpha$ and $\alpha^*$ in the
integrand, the practical evaluation of the integral in (3.12) is not an easy
task. Instead, we use the method of generating function [17,31] which allows us
to transform (3.12) into a multiple-differential form. When we define a
generating function $F(x,y,t)$ by the integral
$$F(x,y,t)=\int P(\alpha,\alpha^*,t)\exp(-\vert\alpha\vert^2+x\alpha+y\alpha^*)
d^2\alpha,\eqno(3.13)$$
we see that the density matrix is related to the generating function by
$$<m\vert\rho(t)\vert n>={1\over\sqrt{m!n!}}({\partial\over\partial x})^m
({\partial\over\partial y})^nF(x,y,t)\vert_{x=y=0}.\eqno(3.14)$$
Since the $P$ representation is in a Gaussian form, the right-hand side of
(3.13) can be evaluated analitically by making use of the identity
$$\int\exp(-a\vert z\vert^2+bz+cz^*+ez^2+fz^{*2})d^2z={\pi\over\sqrt{a^2-4ef}}
\exp{abc+b^2f+c^2e\over a^2-4ef},\eqno(3.15)$$
which is convergent for ${\rm Re}~a>\vert e^*+f\vert$, while $b,c$ may be
arbitrary. We find:
$$F={2\over \sqrt {\vert A\vert}}\exp\{xy-{1\over A}[\sigma_{11}(x-\bar
\alpha_0^*)^2+\sigma_{22}(y-\bar\alpha_0)^2-2(\sigma_{12}+2)(x-\bar\alpha_0^*)
(y-\bar\alpha_0)]\},\eqno(3.16)$$
where $A\equiv d-4(\sigma_{12}+1), ~~d\equiv {\rm det}\sigma=\sigma_{11}\sigma
_{22}-\sigma_{12}^2.$
A formula for the density matrix can be written down by applying the relation
(3.14) to the generating function (3.16). We get
$$<m\vert\rho(t)\vert n>=2{\sqrt{m!n!}\over \sqrt {\vert A\vert}}\exp\{[-
{1\over A}[\sigma_{22}\bar\alpha_0^2+\sigma_{11}\bar\alpha_0^{*2}-2(\sigma_{12}
+2)\vert\bar\alpha_0\vert^2]\}$$
$$\times\sum_{n_1,n_2,n_3=0}{(-1)^{n_1+n_2}2^{m+n-2(n_1+n_2+n_3)}\over n_1!n_2!
n_3!(m-2n_1-n_3)!(n-2n_2-n_3)!}E,\eqno(3.17)$$
where
$$E={\sigma_{11}^{n_1}\sigma_{22}^{n_2}(d-2\sigma_{12})^{n_3}[\sigma_{11}
\bar\alpha_0^*-(\sigma_{12}+2)\bar\alpha_0]^{m-2n_1-n_3}[\sigma_{22}\bar
\alpha_0-2(\sigma_{12}+2)\bar\alpha_0^*]^{n-2n_2-n_3}\over A^{m+n-(n_1+n_2+n_3)
}}.$$
The expression (3.17) is the density matrix corresponding to the initial
coherent state. At time $t=0,$ the functions $\sigma_{11},$ $\sigma_{22}$ and
$\sigma_{12}$ vanish and $\bar\alpha_0$ goes to $\alpha_0.$ In this case the
density matrix reduces to
$$<m\vert\rho(0)\vert n>={1\over\sqrt{m!n!}}\alpha_0^n\alpha_0^{*m}\exp(-\vert
\alpha_0\vert^2),\eqno(3.18) $$
which is the initial Glauber packet. For the diagonal case the initial density
matrix becomes the Poisson distribution. At infinity of time, the density
matrix (3.17) goes to the Bose-Einstein distribution
$$<m\vert\rho(\infty)\vert n>={<n>^m\over(1+<n>)^{m+1}}\delta_{mn}.\eqno(3.19)
$$
For other specific initial conditions, it is more convenient to proceed to
solve directly the master equation in order to extract a closed form for the
density matrix [17].

{\it 3.2 Wigner distribution function}

The Fokker-Planck equation (3.1) can also be written in terms of real
coordinates $x_1$ and $x_2$ (or the averaged position and momentum coordinates
of the harmonic oscillator) defined by $\alpha=x_1+ix_2,~\alpha^*=x_1-ix_2,$
as follows [16]:
$${\partial W\over\partial t}=\sum_{i,j=1,2}A_{ij}{\partial\over
\partial x_i}(x_jW)+{1\over 2}\sum_{i,j=1,2}Q^W_{ij}{\partial^2\over
\partial x_i\partial x_j}W,\eqno(3.20)$$
where
$$A=\left(\matrix{\lambda-\mu&-\omega\cr
\omega&\lambda+\mu\cr}\right),~~~
Q^W={1\over\hbar}\left(\matrix{m\omega D_{qq}&D_{pq}\cr
D_{pq}&{D_{pp}\over m\omega}\cr}\right).\eqno(3.21)$$
Since the drift coefficients are linear in the variables $x_1$ and $x_2$ and
the diffusion coefficients are constant with respect to $x_1$ and $x_2,$ (3.20)
describes an Ornstein-Uhlenbeck process [35,43]. Following the method developed
by Wang and Uhlenbeck [35], we shall solve this Fokker-Planck equation, subject
to either the wave-packet type or the $\delta$-function type of initial
conditions. By changing the variables $x_1$ and $x_2$ of (3.20) via the
relations
$$z_1=ax_1+bx_2,~~ z_2=cx_1+dx_2,\eqno(3.22)$$
the Fokker-Planck equation (3.20) is transformed into the standard linearized
partial differential equation [35]  expressed as
$${\partial W(z_1,z_2,t)\over\partial t}=(-\nu_1{\partial\over\partial z_1}z_1-
\nu_2{\partial\over\partial z_2}z_2+{1\over 2}\sum_{i,j=1,2}D_{ij}
{\partial^2\over\partial z_i\partial z_j})W(z_1,z_2,t).\eqno(3.23)$$
In deriving this equation we have put $a=c^*={\mu-i\Omega\over\omega}, b=d=1,
\nu_1=\nu_2^*\equiv-\lambda-i\Omega$ and
$$D_{11}=D_{22}^*\equiv{1\over\hbar\omega}[(\mu-i\Omega)^2mD_{qq}+2(\mu-i
\Omega)D_{pq}+{D_{pp}\over m}],$$
$$D_{12}=D_{21}\equiv{1\over\hbar}(m\omega D_{qq}+{2\mu\over\omega}D_{pq}+{D_
{pp}\over m\omega}).\eqno(3.24)$$

1) When the Fokker-Planck equation for the coherent state representation is
subject to the initial condition $\delta(\alpha-\alpha_0)\delta(\alpha^*-\alpha
^*_0),$ then the use of the relation between the Wigner distribution function
and $P$ representation
$$W(\alpha,\alpha^*,t)={2\over\pi}\int P(\beta,\beta^*,t)\exp(-2\vert\alpha-
\beta\vert^2)d^2\beta\eqno(3.25)$$
leads to a Gaussian form for the initial Wigner distribution function at $t=0.$
The initial Wigner distribution function can in turn be related to the Wigner
distribution function expressed in terms of $x_{10}$ and $x_{20}$ -- the
initial values of $x_1$ and $x_2$ at $t=0,$ respectively. Then we have the
relation
$$W_w(x_1,x_2,0)=W_w(\alpha,\alpha^*,0)={2\over\pi}\exp(-2\vert
\alpha-\alpha_0\vert^2).\eqno(3.26)$$
The right-hand side of (3.26) can be expressed in terms of real coordinates
$x_1$ and $x_2$ and we thus get the expression which corresponds to the initial
condition of a wave packet [41]. Accordingly, we now look for the solution of
the Fokker-Planck equation (3.20) subject to the initial condition
$$W_w(x_1,x_2,0)={2\over\pi}\exp\{-2[(x_1-x_{10})^2+(x_2-x_{20})^2]\}.\eqno
(3.27)$$
By changing the variables $x_1$ and $x_2$ into $z_1$ and $z_2, (z_1=z^*_2\equiv
z)$, this initial condition is seen to be transformed into
$$W_w(z,z^*,0)={2\over\pi}\exp\{{2\omega^2\over \Omega^2}[q(z-z_0)^2+q^*(z^*-z^
*_0)^2-\vert z-z_0\vert^2]\},\eqno(3.28)$$
where $z_0$ is the initial value of $z$ and $q={\mu(\mu+i\Omega)\over 2\omega^2
}.$ The solution of (3.20) subject to the initial condition (3.28) is found
to be
$$W_w(z,z^*,t)={2\Omega\over \pi\omega\sqrt{\vert B_w\vert}}\exp\{-{1\over 2B_
w}[g_2(z-z_0e^{\nu_1t})^2+g_1(z^*-z_0^*e^{\nu_2t})^2-g_3\vert z-z_0e^{\nu_1t}
\vert ^2]\},\eqno(3.29)$$
where
$$B_w=g_1g_2-{1\over 4}g_3^2, g_1=g_2^*=q^*e^{2\nu_1 t}+{D_{11}\over 2\nu_1}(e^
{2\nu_1t}-1), g_3=e^{-2\lambda t}+{D_{12}\over\lambda}(1-e^{-2\lambda t}).
\eqno(3.30)$$
In terms of real variables $x_1$ and $x_2$ we have:
$$W_w(x_1,x_2,t)={1\over \pi\sqrt{\vert B_w\vert}}\exp\{-{1\over 2B_w}[\phi_w
(x_1-\bar x_1)^2+\psi_w(x_2-\bar x_2)^2+\chi_w(x_1-\bar x_1)(x_2-\bar x_2)]\},
\eqno(3.31)$$
where
$$\phi_w=g_1a^{*2}+g_2a^2-g_3,~\psi_w=g_1+g_2-g_3,~\chi_w=2(g_1a^*+g_2a)-g_3(a
+a^*).\eqno(3.32)$$
The functions $\bar x_1$ and $\bar x_2$, which are also oscillating functions,
are given by
$$\bar x_1=e^{-\lambda t}[x_{10}(\cos\Omega t+{\mu\over\Omega}\sin\Omega t)+
x_{20}{\omega\over\Omega}\sin\Omega t],$$
$$\bar x_2=e^{-\lambda t}[x_{20}(\cos\Omega t-{\mu\over\Omega}\sin\Omega t)-
x_{10}{\omega\over\Omega}\sin\Omega t].\eqno(3.33)$$

2) If the Fokker-Planck equation (3.20) is subject to the $\delta$-function
type of initial condition, the Wigner distribution function is given by
$$W(z,z^*,t)={2\Omega\over\pi\omega\sqrt{\vert B\vert}}\exp\{-{1\over 2B}[f_2(
z-z_0e^{\nu_1t})^2+f_1(z^*-z_0^*e^{\nu_2t})^2-2f_3\vert z-z_0e^{\nu_1t}\vert^2
]\},\eqno(3.34)$$
where
$$B=f_1f_2-f_3^2,~~f_1=f_2^*={D_{11}\over 2\nu_1}(e^{2\nu_1t}-1),~~
f_3={D_{12}\over 2\lambda}(1-e^{-2\lambda t}).\eqno(3.35)$$
In terms of real variables $x_1$ and $x_2$ we have:
$$W(x_1,x_2,t)={1\over 2\pi\sqrt{\vert B\vert}}\exp\{-{1\over 2B}[\phi_d
(x_1-\bar x_1)^2+\psi_d(x_2-\bar x_2)^2+\chi_d(x_1-\bar x_1)(x_2-\bar x_2)]\},
\eqno(3.36)$$
where
$$\phi_d=f_1a^{*2}+f_2a^2-f_3,~\psi_d=f_1+f_2-f_3,~\chi_d=2(f_1a^*+f_2a)-f_3(a
+a^*).\eqno(3.37)$$
So, one gets a 2-dimensional Gaussian distribution with the average values
$\bar x_1$ and $\bar x_2$ and the variances $\phi_d,$$\psi_d$ and $\chi_d.$
Obviously, at time $t=0,$ all functions $B, f_1,f_2,f_3$ vanish and the
Wigner distribution function goes to the $\delta$-function.

When time $t$ goes to infinity, $\bar x_1$ and $\bar x_2$ vanish and we obtain
the steady state solution:
$$W(x_1,x_2)={1\over 2\pi\sqrt{{\rm det}\sigma^W(\infty)}}\exp[-{1\over 2}
\sum_{i,j=1,2}(\sigma^W)^{-1}_{ij}(\infty)x_ix_j].\eqno(3.38)$$
The stationary covariance matrix $\sigma^W(\infty)$ can be determined from the
algebraic equation
$$A\sigma^W(\infty)+\sigma^W(\infty)A^{\rm T}=Q^W.\eqno(3.39)$$
We obtain:
$$\sigma_{11}^W(\infty)={(2\lambda(\lambda+\mu)+\omega^2)Q_{11}^W+\omega^2Q_
{22}^W+2\omega(\lambda+\mu)Q_{12}^W\over 4\lambda(\lambda^2+\omega^2-\mu^2)},$$
$$\sigma_{22}^W(\infty)={\omega^2Q_{11}^W+(2\lambda(\lambda-\mu)+\omega^2)Q_{22
}^W-2\omega(\lambda-\mu)Q_{12}^W\over 4\lambda(\lambda^2+\omega^2-\mu^2)},
\eqno(3.40)$$
$$\sigma_{12}^W(\infty)={-\omega(\lambda+\mu)Q_{11}^W+\omega(\lambda-\mu)Q_
{22}^W+2(\lambda^2-\mu^2)Q_{12}^W\over 4\lambda(\lambda^2+\omega^2-\mu^2)}.$$

{\bf 4. Conclusions}

The Lindblad theory provides a selfconsistent
treatment of damping as a possible extension of quantum mechanics to open
systems. In the present paper we have studied the one-dimensional harmonic
oscillator with dissipation within the framework of this theory. We have
carried out an extended calculation of the expectation values of various
dynamical operators involved in the master equation, especially the first two
moments and the density matrix. Generally, the time evolution of the density
matrix as well as the expectation values of dynamical operators are
characterized by complex functions with an oscillating element $\exp(\pm
i\sqrt{\omega^2-\mu^2}t)$ multiplied by the damping factor $\exp(-\lambda t).$
We deduced the density matrix from the solution of the Fokker-Planck equation
for the coherent state representation, obtained from the master equation for
the density operator. For a thermal bath, when the asymptotic state is a Gibbs
state, a Bose-Einstein distribution results as density matrix. Also for the
case that the initial density matrix is chosen as a Glauber packet, a simple
analytical expression for the density matrix has been derived. The density
matrix can be used in various physical applications where a Bosonic degree of
freedom moving in a harmonic oscillator potential is damped. For example, one
needs to determine nondiagonal transition elements of the density matrix, if
an oscillator is perturbed by a weak electromagnetic field in addition to its
coupling to a heat bath. From the master equation of the damped quantum
oscillator we have also derived the corresponding Fokker-Planck equation in the
Wigner $W$ representation. The Fokker-Planck equation which we obtained
describes an Ornstein-Uhlenbeck process. We have solved this equation by using
the Wang-Uhlenbeck method of transforming it into a linearized partial
differential equation for the Wigner function, subject to either the Gaussian
type or the $\delta$-function type of initial conditions and showed that the
Wigner functions are two-dimensional Gaussians with different widths.

{\bf References}

\item{ 1.}
R. W. Hasse, J. Math. Phys. {\bf 16} (1975) 2005

\item{ 2.}
E. B. Davies, Quantum Theory of Open Systems (Academic Press, New York, 1976)

\item{ 3.}
H. Spohn, Rev. Mod. Phys. {\bf 52} (1980) 569

\item{ 4.}
H. Dekker, Phys. Rep. {\bf 80} (1981) 1

\item{ 5.}
P. Exner, Open Quantum Systems and Feynman Integrals (Reidel, Dordrecht,1985)

\item{ 6.}
K. H. Li, Phys. Rep. {\bf 134} (1986) 1

\item{ 7.}
G. Lindblad, Commun. Math. Phys. {\bf 48} (1976) 119

\item{ 8.}
G. Lindblad, Rep. Math. Phys. {\bf 10} (1976) 393

\item{ 9.}
G. Lindblad, Non-Equilibrium Entropy and Irreversibility (Reidel, Dordrecht,
1983)

\item{10.}
A. Sandulescu and H. Scutaru, Ann. Phys. (N.Y.) {\bf 173} (1987) 277

\item{11.}
A. Sandulescu, H. Scutaru and W. Scheid, J. Phys. A - Math. Gen. {\bf 20}
(1987) 2121

\item{12.}
A. Pop, A. Sandulescu, H. Scutaru and W. Greiner, Z. Phys. A {\bf 329} (1988)
357

\item{13.}
A. Isar, A. Sandulescu and W. Scheid, Int. J. Mod. Phys. A {\bf 5}
(1990) 1773

\item{14.}
A. Isar, A. Sandulescu and W. Scheid, J. Phys. G - Nucl. Part. Phys. {\bf 17}
(1991) 385

\item{15.}
A. Sandulescu and E. Stefanescu, Physica A {\bf 161} (1989) 525

\item{16.}
A. Isar, W. Scheid and A. Sandulescu, J. Math. Phys. {\bf 32} (1991) 2128

\item{17.}
A. Isar, A. Sandulescu and W. Scheid, J. Math. Phys. {\bf 34} (1993) 3887

\item{18.}
C. W. Gardiner and M. J. Collet, Phys. Rev. A {\bf 31} (1985) 3761

\item{19.}
T. A. B. Kennedy and D. F. Walls, Phys. Rev. A {\bf 37} (1988) 152

\item{20.}
C. M. Savage and D. F. Walls, Phys. Rev. A {\bf 32} (1985) 2316

\item{21.}
G. S. Agarwal, Phys. Rev. {\bf 178} (1969) 2025

\item{22.}
G. S. Agarwal, Phys. Rev. A {\bf 4} (1971) 739

\item{23.}
S. Dattagupta, Phys. Rev. A {\bf 30} (1984) 1525

\item{24.}
N. Lu, S. Y. Zhu and G. S. Agarwal, Phys. Rev. A {\bf 40} (1989) 258

\item{25.}
S. Carrusotto, Phys. Rev. A {\bf 11} (1975) 1397

\item{26.}
H.Haken, Handbuch der Physik, vol.XXV/2C (Springer, Berlin, 1970) p.29

\item{27.}
W. H. Louisell, Quantum Statistical Properties of Radiation (Wiley, New York,
1973)

\item{28.}
R. Graham, Springer Tracts in Mod. Phys. {\bf 66} (Springer, Berlin, 1973) p.1

\item{29.}
F.Haake, Springer Tracts in Mod. Phys. {\bf 66} (Springer, Berlin, 1973) p.98

\item{30.}
C. W. Gardiner, Handbook of Stochastic Methods (Springer, Berlin, 1982)

\item{31.}
S. Jang, Nucl. Phys. A {\bf 499} (1989) 250

\item{32.}
H. Hofmann and P. J. Siemens, Nucl. Phys. A {\bf 275} (1977) 464

\item{33.}
H. Hofmann, C. Gr\'egoire, R. Lucas and C. Ng\^o, Z. Phys. A  {\bf 293} (1979)
229

\item{34.}
E. M. Spina and H. A. Weidenm\"uller, Nucl. Phys. A {\bf 425} (1984) 354

\item{35.}
M. C. Wang and G. E. Uhlenbeck, Rev. Mod. Phys. {\bf 17} (1945) 323

\item{36.}
K. E. Cahill, Phys. Rev. {\bf 180} (1969) 1239

\item{37.}
K. E. Cahill, Phys. Rev. {\bf 180} (1969) 1244

\item{38.}
H. Haken, Rev. Mod. Phys. {\bf 47} (1975) 67

\item{39.}
H. Risken, The Fokker-Planck Equation (Springer, Berlin, 1984)

\item{40.}
P. D. Drummond and C. W. Gardiner, J. Phys. A - Math. Gen. {\bf 13} (1980) 2353

\item{41.}
E. J. Glauber, Phys. Rev. {\bf 131} (1963) 2766

\item{42.}
M. Hillery, R.F. O 'Connell, M.O. Scully and E.P. Wigner, Phys.Rep.
{\bf 106} (1984) 121

\item{43.}
G. E. Uhlenbeck and L. S. Ornstein, Phys. Rev. {\bf 36} (1930) 823
\vfill
\eject
\bye